\documentstyle[11pt,epsfig]{article} \textheight 700pt \textwidth
480pt \oddsidemargin 0pt \voffset -2.5cm
\title{\bf Bianchi-I classical and quantum spinor cosmology with signature change}
\author{B. Vakili$^1$\thanks{email:
b-vakili@cc.sbu.ac.ir}
  and H. R. Sepangi$^{1,2}$\thanks{email:
hr-sepangi@cc.sbu.ac.ir}
\\ $^1${\small Department of Physics, Shahid Beheshti University, Evin, Tehran 19839, Iran}\\$^2${\small
Institute for Studies in Theoretical Physics and Mathematics, P.O.
Box 19395-5746, Tehran, Iran }}
\begin{document}
\maketitle 

\begin{abstract}
We study the classical and quantum evolution of a universe in
which the matter source is a massive Dirac spinor field and the
universe is described by a Bianchi type I metric. We focus
attention on the those classical solutions that admit a degenerate
metric in which the scale factors have smooth behavior in
transition from a Euclidean to a Lorentzian domain and show that
this transition happens when the cosmological constant, $\Lambda$,
is negative. The resulting quantum cosmology and the corresponding
Wheeler-DeWitt equation are also studied and closed form
expressions for the wave functions of the universe is presented.
We have shown that there is a close relationship between the
quantum states and signature changing classical solutions,
suggesting a mechanism for creation of a Lorentzian universe from
a Euclidean region by a continuous change of signature. The
quantum solutions
also represent a quantization rule for the mass of the spinor field.\vspace{10mm}
\end{abstract}\pagebreak
\section{Introduction}
A question of great importance in cosmology is that of the initial
conditions from which the universe has evolved. As is well known,
standard cosmological models based on classical general relativity
have no convincingly precise answer to this question. This can be
traced to the fact that these models suffer from the presence of
an initial singularity, the so-called ``Big-Bang'' singularity.
Any hope of dealing with such singularities would be in the
development of a concomitant and conducive quantum theory of
gravity. In the absence of a full theory of quantum gravity, it
would be useful to describe the quantum state of the universe
within the context of quantum cosmology, introduced in the works
of DeWitt \cite{1}. In this formalism which is based on the
canonical quantization procedure, the evolution of universe is
described by a wave function in the mini-superspace. Two major
approaches in this scenario are the {\it tunneling} proposal first
developed by Vilenkin \cite{2}-\cite{6}, and the {\it no-boundary}
proposal of Hartle and Hawking \cite{7}-\cite{9}. In the tunneling
proposal the wave function is so constructed as to create a
universe emerging from {\it nothing} by a tunneling procedure
through a potential barrier in the sense of usual quantum
mechanics. In the \textit{no-boundary} proposal of Hartle and
hawking on the other hand, the wave function is constructed by a
path integral over all compact Euclidean 4-manifolds. A problem
related to this approach is that of signature transition from a
Euclidean to a Lorentzian manifold. The notion of signature
transition was first addressed in \cite{7}, where the authors
argued that in quantum cosmology amplitudes for gravity should be
expressed as the sum of all compact Riemannian manifolds whose
boundaries are located at the signature changing hypersurface. In
the traditional point of view, a feature in general relativity is
that one usually fixes the signature of the space-time metric
before trying to solve Einstein's field equations. However there
is no {\it a priori} reason for doing so and it is well known that
the field equations do not demand this property, that is, if one
relaxes this condition one may find solutions to the field
equations which when parameterized suitably, can either have
Euclidean or Lorentzian signature \cite{10}-\cite{16}.

An important ingredient in any model theory related to cosmology
is the choice of the matter field used to couple with gravity. The
most widely used matter source has traditionally been the perfect
fluid. However, the scalar field has also been playing an
increasingly important role in the more resent cosmological
models. One main reason is that being a scalar field makes it
somewhat easy to work with. Another matter field which has
occasionally been studied in the literature is the massless or
massive spinor field as the source of gravity. In general,
theories studying spinor fields coupled to gravity result in
Einstein-Dirac systems which are not easy to solve. The
cosmological solutions of such systems have been studied in a few
cases by a number of authors \cite{16}-\cite{20}. The references
\cite{16} and \cite{17} are notable in that the quantization of a
spinor field coupled to gravity is studied in a Robertson-Walker
background.

In this paper we deal with the classical and quantum cosmology of
a model in which a massive self interacting spinor field is
coupled to gravity in a Bianchi type-I space-time. Bianchi models
are the most popular anisotropic and homogeneous cosmologies and
are studied in different levels by many authors,
\cite{19}-\cite{26}. From the classical solutions of the resulting
Einstein-Dirac system we have chosen those that exhibit a smooth
transition from a Euclidean to a Lorentzian region, {\it i.e.},
signature changing solutions. Although in classical gravity such
solutions may not seem to be too interesting, we show that there
is a close relationship between these solutions and the those
resulting from the Wheeler-DeWitt (WD) equation in the
corresponding quantum cosmology. It turns out that the WD equation
possesses exact solutions in terms of confluent hypergeometric
functions. We show that these solutions predict creation of a
universe by a continuous transition from a classically forbidden
(Euclidean) to a classically allowed (Lorentzian) domain and that
they are in agreement with the classical signature changing
solutions. We also show that imposing boundary conditions on the
wave function of the universe suggests a quantization condition
for the mass of the spinor field.

\section{Preliminary setup}
The Bianchi type I space-time is characterized in a comoving
coordinate system by the metric
\begin{equation}\label{A}
ds^2=-dt^2+a^2(t)dx^2+b^2(t)dy^2+c^2(t)dz^2,
\end{equation}
where $a(t)$, $b(t)$ and $c(t)$ are scale factors in $x$, $y$ and
$z$ directions respectively. This metric is the simplest
anisotropic and homogeneous cosmological model which, upon having
equal scale factors becomes the flat Robertson-Walker metric. Such
space-times have an Abelian symmetry group of translations with
Killing vector fields ${\bf
\xi}=(\partial_x,\partial_y,\partial_z)$. Of course, all the
structure constants of such a symmetry group are zero. The scalar
curvature corresponding to metric (\ref{A}) is
\begin{equation}\label{B}
{\cal
R}=2\left(\frac{\ddot{a}}{a}+\frac{\ddot{b}}{b}+\frac{\ddot{c}}{c}+
\frac{\dot{a}\dot{b}}{ab}+\frac{\dot{b}\dot{c}}{bc}+\frac{\dot{c}\dot{a}}{ca}\right),
\end{equation}
where a dot represents differentiation with respect to $t$. We may
parameterize the metric in such a way as to allow the Euclidean
signature $(+,+,+,+)$ becoming  Lorentzian  $(-,+,+,+)$
\cite{11}-\cite{15}. To this and other ends, we parameterize the
metric as in \cite{14} and \cite{15} by adapting the chart
$\{\beta,x,y,z\}$ where the hypersurface of signature change would
be characterized by $\beta=0$. The metric can then be
parameterized in terms of the scale factors $a(\beta)$,
$b(\beta)$, $c(\beta)$ and laps function $\beta$ to take the form
\begin{equation}\label{C}
ds^2=-\beta d\beta^2+a^2(\beta)dx^2+b^2(\beta)dy^2+c^2(\beta)dz^2.
\end{equation}
It is now clear that the sign of the laps function $\beta$
determines the signature of the metric, being Lorentzian if
$\beta>0$ and Euclidean if $\beta<0$. For the Lorentzian region
the traditional cosmic time can be recovered by the substitution
$t=\frac{2}{3}\beta^{3/2}$. Also,  adapting the chart
$\{t,x,y,z\}$ in this region we shall write any dynamical field
such as the scale factors or matter fields as
$\Phi(t)=\Phi(\beta(t))$. As a result of the above discussion, we
see that the signature changing hypersurface divides the manifold
into two domains, Euclidean ${\cal M_E}$ and Lorentzian ${\cal
M_L}$. As has been discussed in \cite{13}, one may write
\begin{equation}\label{D}
{\cal M_E}\cap{\cal M_L}=\emptyset,\hspace{.5cm} \bar{{\cal
M_E}}\cap \bar{{\cal M_L}}=\Sigma, \hspace{.5cm} \bar{{\cal
M_E}}\cup \bar{{\cal M_L}}={\cal M},
\end{equation}
where $\Sigma$ represents the signature changing hypersurface,
that is the hypersurface with $\beta=0$ and ${\cal M}$ is the
total manifold. From the point of view of the Einstein field
equations, Euclidean and Lorentzian regions are classically
forbidden and allowed solutions of the gravitational field
equations respectively. We formulate our differential equations in
a region which does not include $\beta=0$ and seek solutions for
any dynamical field that smoothly passes through the $\beta=0$
hypersurface. These solutions are called signature changing
solutions and we shall see that they are classical description of
the quantum cosmological states of the model. Indeed, we will
encounter the Euclidean and Lorentzian regions again when dealing
with the solutions of the WD equation later on.

To construct the field equations, let us start with the action
\begin{equation}\label{E}
{\cal S}=\int (L_{grav}+L_{matt})\sqrt{-g}d^4x,
\end{equation}
where
\begin{equation}\label{F}
L_{grav}={\cal R}-\Lambda,
\end{equation}
is the Einstein-Hilbert Lagrangian for the gravitational field
with cosmological constant $\Lambda$, and $L_{matt}$ represents
the Lagrangian of the matter source, which we assume to be a
massive spinor field. As is well known, the Dirac equation
describing dynamics of a spinor field $\psi$ can be obtained from
the Lagrangian
\begin{equation}\label{G}
L_{matt}=\frac{1}{2}\left[\bar{\psi}\gamma^{\mu}(\partial_{\mu}+
\Gamma_{\mu})\psi-\bar{\psi}(\overleftarrow{\partial_{\mu}}-\Gamma_{\mu})\gamma^\mu\psi
\right]-V(\bar{\psi},\psi),
\end{equation}
where $\gamma^{\mu}$ are the Dirac matrices associated with the
space-time metric satisfying the Clifford algebra
$\{\gamma^{\mu},\gamma^{\nu}\}=2g^{\mu \nu}$, $\Gamma_{\mu}$ are
spin connections and $V(\bar{\psi},\psi)$ is a potential
describing the interaction of the spinor field with itself. The
$\gamma^{\mu}$ matrices are related to the flat Dirac matrices,
$\gamma^a$, through the tetrads $e^a_{\mu}$ as follows
\begin{equation}\label{H}
\gamma^{\mu}=e^{\mu}_{a}\gamma^a, \hspace{.5cm}
\gamma_{\mu}=e^a_{\mu}\gamma_a.
\end{equation}
For metric (\ref{A}) the tetrads can be easily obtained from their
definition, that is, $g_{\mu \nu}=e^a_{\mu}e^b_{\nu}\eta_{ab}$,
leading to
\begin{equation}\label{I}
e^a_{\mu}=\mbox{diag}(1,a,b,c),\hspace{.5cm}
e^{\mu}_a=\mbox{diag}(1,1/a,1/b,1/c).
\end{equation}
Also, the spin connections satisfy the relation
\begin{equation}\label{J}
\Gamma_{\mu}=\frac{1}{4}g_{\nu\lambda}(\partial_{\mu}e^{\lambda}_a+
\Gamma^{\lambda}_{\sigma\mu}e^{\sigma}_a)\gamma^{\nu}\gamma^a.
\end{equation}
Thus, for the line element (\ref{A}), use of (\ref{H}) and
(\ref{I}) yields
\begin{equation}\label{K}
\Gamma_0=0,\hspace{.5cm}\Gamma_1=-\frac{\dot{a}}{2}\gamma^0
\gamma^1,\hspace{.5cm} \Gamma_2=-\frac{\dot{b}}{2}\gamma^0
\gamma^2,\hspace{.5cm} \Gamma_3=-\frac{\dot{c}}{2}\gamma^0
\gamma^3.
\end{equation}
Here, $\gamma^0$ and $\gamma^i$ are the Dirac matrices in
Minkowski space and we have adopted the following representation
\cite{27}
\begin{equation}\label{L}
\gamma^0=\left(%
\begin{array}{cc}
  -i & 0 \\
  0 & i \\
\end{array}%
\right),\hspace{.5cm} \gamma^i=\left(%
\begin{array}{cc}
  0 & \sigma^i \\
  \sigma^i & 0 \\
\end{array}%
\right).
\end{equation}
The final remark about the Lagrangian (\ref{G}) is that
consistency of Einstein field equations with a spinor field as the
matter source in the background metric (\ref{A}) requires the
spinor field $\psi$ to be dependent on $t$ only, that is
$\psi=\psi(t)$ \cite{16}.

The preliminary set-up for writing the action is now complete. By
substituting (\ref{B}), (\ref{F}) and (\ref{G}) into (\ref{E}) and
integrating over spatial dimensions, we are led to an effective
Lagrangian in the mini-superspace $\{a,b,c,\psi,\bar{\psi}\}$
\begin{equation}\label{M}
{\cal L}=\dot{a}\dot{b}c+a\dot{b}\dot{c}+\dot{a}b\dot{c}+\Lambda
abc+\frac{1}{2}abc[\bar{\psi}\gamma^0
\dot{\psi}-\dot{\bar{\psi}}\gamma^0 \psi-2V(\bar{\psi},\psi)].
\end{equation}
\section{Field equations}
Variation of Lagrangian (\ref{M}) with respect to $\bar{\psi}$,
$\psi$, $a$, $b$ and $c$ yields the equations of motion for the
spinor and gravitational fields respectively, that is
\begin{equation}\label{N}
\dot{\psi}+\frac{1}{2}\left(\frac{\dot{a}}{a}+\frac{\dot{b}}{b}+
\frac{\dot{c}}{c}\right)\psi+\gamma^0\frac{\partial
V}{\partial\bar{\psi}}=0,
\end{equation}
\begin{equation}\label{1}
\dot{\bar{\psi}}+\left(\frac{\dot{a}}{a}+\frac{\dot{b}}{b}+
\frac{\dot{c}}{c}\right)\bar{\psi}-\frac{\partial
V}{\partial \psi}\gamma^0=0,
\end{equation}

\begin{equation}\label{O}
\frac{\ddot{b}}{b}+\frac{\ddot{c}}{c}+\frac{\dot{b}\dot{c}}{bc}-
\Lambda=\frac{1}{2}\left[\bar{\psi}\frac{\partial
V}{\partial \bar{\psi}}+\frac{\partial V}{\partial \psi}\psi
\right]-V(\bar{\psi},\psi),
\end{equation}

\begin{equation}\label{2}
\frac{\ddot{c}}{c}+\frac{\ddot{a}}{a}+\frac{\dot{a}\dot{c}}{ac}-
\Lambda=\frac{1}{2}\left[\bar{\psi}\frac{\partial
V}{\partial \bar{\psi}}+\frac{\partial V}{\partial \psi}\psi
\right]-V(\bar{\psi},\psi),
\end{equation}

\begin{equation}\label{3}
\frac{\ddot{a}}{a}+\frac{\ddot{b}}{b}+\frac{\dot{a}\dot{b}}{ab}-
\Lambda=\frac{1}{2}\left[\bar{\psi}\frac{\partial
V}{\partial \bar{\psi}}+\frac{\partial V}{\partial \psi}\psi
\right]-V(\bar{\psi},\psi).
\end{equation}
Also, we have the ``zero- energy'' condition given by
\begin{equation}\label{Q}
{\cal H}=\frac{\partial {\cal L}}{\partial
\dot{a}}\dot{a}+\frac{\partial {\cal L}}{\partial
\dot{b}}\dot{b}+\frac{\partial {\cal L}}{\partial
\dot{c}}\dot{c}+\frac{\partial {\cal L}}{\partial
\dot{\psi}}\dot{\psi}+\dot{\bar{\psi}}\frac{\partial {\cal
L}}{\partial \dot{\bar{\psi}}}-{\cal L}=0,
\end{equation}
which yields the constraint equation
\begin{equation}\label{R}
\frac{\dot{a}\dot{b}}{ab}+\frac{\dot{b}\dot{c}}{bc}+
\frac{\dot{a}\dot{c}}{ac}-\Lambda=-V(\bar{\psi},\psi).
\end{equation}
It is clear that the right hand side of equations
(\ref{O})-(\ref{3}) and (\ref{R}) represents the components of the
energy-momentum tensor where the energy density of the spinor
field is given by
\begin{equation}\label{S}
\rho=T_{00}=-V(\bar{\psi},\psi).
\end{equation}
Finally we write the Hamiltonian constraint (\ref{Q}) in terms of
the momenta conjugate to our dynamical variables, giving
\begin{equation}\label{T}
p_a=\frac{\partial {\cal L}}{\partial
\dot{a}}=\dot{b}c+b\dot{c},\hspace{.5cm} p_b=\frac{\partial {\cal
L}}{\partial \dot{b}}=a\dot{c}+\dot{a}c,\hspace{.5cm}
p_c=\frac{\partial {\cal L}}{\partial \dot{c}}=a\dot{b}+\dot{a}b,
\end{equation}
and also
\begin{equation}\label{U}
p_{\psi}=\frac{\partial {\cal L}}{\partial
\dot{\psi}}=\frac{1}{2}abc\bar{\psi}\gamma^0,\hspace{.5cm}
p_{\bar{\psi}}=\frac{\partial {\cal L}}{\partial
\dot{\bar{\psi}}}=-\frac{1}{2}abc \gamma^{0}\psi.
\end{equation}
In terms of conjugate momenta, the Hamiltonian is given by
\begin{equation}\label{V}
{\cal H}=-\frac{1}{4}\left(\frac{a}{bc}p_a^2 +\frac{b}{ac}p_b^2
+\frac{c}{ab}p_c^2\right)+\frac{1}{2}\left(\frac{p_a
p_b}{c}+\frac{p_b p_c}{a}+\frac{p_a p_c}{b}\right)-\Lambda
abc+abcV(\bar{\psi},\psi)=0.
\end{equation}
In the next section we shall present the solutions of the
classical field equations (\ref{N})-(\ref{3}). Although, these
equations can be solved after a suitable form for the potential
$V(\bar{\psi},\psi)$ has been chosen, see \cite{19} and \cite{20},
Hamiltonian (\ref{V}) has not the desired form for the
construction of the WD equation describing the relevant quantum
cosmology. Thus, to simplify Lagrangian (\ref{M}), consider the
following change of variables \cite{28}
\begin{equation}\label{W}
a=e^{u+v+\sqrt{3}w},\hspace{.5cm}
b=e^{u+v-\sqrt{3}w},\hspace{.5cm} c=e^{u-2v}.
\end{equation}
In terms of these variables, Lagrangian (\ref{M}) takes the form
\begin{equation}\label{X}
{\cal L}=3\left(\dot{u}^2 -\dot{v}^2
-\dot{w}^2\right)e^{3u}+\Lambda
e^{3u}+\frac{1}{2}e^{3u}\left[\bar{\psi}\gamma^0
\dot{\psi}-\dot{\bar{\psi}}\gamma^0
\psi-2V(\bar{\psi},\psi)\right],
\end{equation}
with the corresponding ``zero-energy'' condition
\begin{equation}\label{Y}
e^{-3u}{\cal
H}=3\left(\dot{u}^2-\dot{v}^2-\dot{w}^2\right)-\Lambda+V(\bar{\psi},\psi)=0.
\end{equation}
Also, the momenta conjugate to $u$, $v$ and $w$ are
\begin{equation}\label{Z}
p_u=\frac{\partial {\cal L}}{\partial
\dot{u}}=6\dot{u}e^{3u},\hspace{.5cm} p_v=\frac{\partial {\cal
L}}{\partial \dot{v}}=-6\dot{v}e^{3u},\hspace{.5cm}
p_w=\frac{\partial {\cal L}}{\partial \dot{w}}=-6\dot{w}e^{3u},
\end{equation}
which give rise to the following Hamiltonian for our dynamical
system
\begin{equation}\label{AB}
{\cal H}=\frac{1}{12}e^{-3u}\left(p_u^2 -p_v^2 -p_w^2
\right)+\left[V(\bar{\psi},\psi)-\Lambda\right]e^{3u}=0.
\end{equation}
Now, variation of Lagrangian (\ref{X}) with respect to its
dynamical variables yields the following field equations
\begin{equation}\label{AC}
\dot{\psi}+\frac{3}{2}\dot{u}\psi+\gamma^0 \frac{\partial
V}{\partial \bar{\psi}}=0,
\end{equation}
\begin{equation}\label{4}
\dot{\bar{\psi}}+\frac{3}{2}\dot{u}\bar{\psi}-\frac{\partial
V}{\partial \psi}\gamma^0=0,
\end{equation}
\begin{equation}\label{AD}
2\ddot{u}+3\dot{u}^2+3(\dot{v}^2+\dot{w}^2)-\Lambda-\frac{1}{2}\left[\bar{\psi}\gamma^0
\dot{\psi}-\dot{\bar{\psi}}\gamma^0
\psi-2V(\bar{\psi},\psi)\right]=0,
\end{equation}
\begin{equation}\label{AA}
\left(\dot{v}e^{3u}\right)^{.}=0,
\end{equation}
\begin{equation}\label{AAA}
\left(\dot{w}e^{3u}\right)^{.}=0.
\end{equation}
Our goal would now be to concentrate on the solutions of these
equations for a suitable form of the potential $V$.
\section{Classical solutions}
The system of differential equations (\ref{AC})-(\ref{AAA}) are
the Einstein-Dirac system for the gravitational field coupled to a
massive self interacting spinor field. Since integrability of this
system directly depends on the choice of a form for
$V(\bar{\psi},\psi)$, it is appropriate to concentrate on this
point first. Interesting forms for such potentials  should involve
terms which would describe self-interacting spinor fields. The
potential is usually an invariant function constructed from the
spinor $\psi$ and its adjoint $\bar{\psi}$ and we require it to
satisfy the following property
\begin{equation}\label{AE}
\bar{\psi}\gamma^0 \frac{\partial V}{\partial
\bar{\psi}}-\frac{\partial V}{\partial \psi}\gamma^0 \psi=0.
\end{equation}
This property is a mild restriction on the potential as may easily
be verified for the most commonly used ones.  One such form for
the potential that has property (\ref{AE}) is
\begin{equation}\label{AF}
V(\bar{\psi},\psi)=m\bar{\psi}\psi+\lambda(\bar{\psi}\psi)^2,
\end{equation}
where $m$ is the mass of the spinor field and $\lambda$ is a
coupling constant. With this potential we can immediately
integrate equations (\ref{AC}) and (\ref{4}) to obtain
$\bar{\psi}\psi={\cal C}e^{-3u}$, where ${\cal C}$ is an
integrating constant. Bearing in the mind from equation (\ref{S})
that the energy density of the spinor field should have a positive
value and with the use of (\ref{AF}) we are led to a negative
value for ${\cal C}$ which we take choose as $-1$,  hence
\begin{equation}\label{AG}
\bar{\psi}\psi=-e^{-3u}.
\end{equation}
The next step is to solve equations (\ref{Y}) and
(\ref{AD})-(\ref{AAA}). With potential (\ref{AF}) these equations
read
\begin{equation}\label{AH}
\dot{v}=c_1e^{-3u},
\end{equation}
\begin{equation}\label{AI}
\dot{w}=c_{2}e^{-3u},
\end{equation}
\begin{equation}\label{AJ}
2\ddot{u}+3\dot{u}^2+\frac{C^2}{3}e^{-6u}-\Lambda=0,
\end{equation}
\begin{equation}\label{AK}
3\dot{u}^2-\frac{C^2}{3}e^{-6u}-\Lambda -me^{-3u}=0,
\end{equation}
where in deriving them we have  integrated equations (\ref{AA})
and (\ref{AAA}) and substituted the results in (\ref{Y}) and
(\ref{AD}). $c_1$ and $c_2$ are two integrating constants and
$C^2=9(c_1^2+c_2^2-\lambda/3)$. Of course, equations (\ref{AJ})
and (\ref{AK}) are not independent and a solution to one should
satisfy  the other automatically. Integration of  equation
(\ref{AK}) would be made easier after a change of variable
$\tau=e^{3u}$, resulting in
\begin{equation}\label{AL}
\dot{\tau}^2=3\Lambda \tau^2+3m\tau+C^2.
\end{equation}
This would now be a simple equation to integrate. Depending  on
the sign of $\Lambda$, there are three classes of solutions for
$\Lambda=0$, $\Lambda<0$ and $\Lambda>0$.  We are interested on
those values which would allow for signature changing solutions to
be constructed. For $\Lambda=0$, we have
\begin{equation}\label{AM}
\tau=\frac{3m}{4}\left(t+A\right)^2-\frac{C^2}{3m},
\end{equation}
where $A$ is a constant. It is clear that this solution, after
substitution $t=\frac{2}{3}\beta^{3/2}$, does not exhibit smooth
transitions from $\beta<0$ to $\beta>0$ regions. In the case of
$\Lambda<0$, integrating equation (\ref{AL}) results in
\begin{equation}\label{AN}
\tau=\frac{m}{-2\Lambda}+\sqrt{\frac{m^2}{4\Lambda^2}+
\frac{C^2}{-3\Lambda}}\cos(\sqrt{-3\Lambda}t),
\end{equation}
where the integrating constant is chosen so that
$\dot{\tau}(0)=0$. The functions $u$, $v$ and $w$ can then be
easily obtained in terms of the evolution parameter $\beta$ as
\begin{equation}\label{AO}
u(\beta)=\ln\left[\frac{m}{-2\Lambda}+\sqrt{\frac{m^2}{4\Lambda^2}+
\frac{C^2}{-3\Lambda}}\cos\left(\frac{2}{3}\sqrt{-3\Lambda}
\beta^{3/2}\right)\right]^{1/3},
\end{equation}
\begin{equation}\label{AP}
v(\beta)=\frac{c_1}{C}\ln\left|\frac{(B-A)\tan\left(\sqrt{-\Lambda/3}\beta^{3/2}\right)+C/\sqrt{-3\Lambda}}
{(B-A)\tan\left(\sqrt{-\Lambda/3}\beta^{3/2}\right)-C/\sqrt{-3\Lambda}}\right|,
\end{equation}
\begin{equation}\label{AQ}
w(\beta)=\frac{c_2}{C}\ln\left|\frac{(B-A)\tan\left(\sqrt{-\Lambda/3}\beta^{3/2}\right)+C/\sqrt{-3\Lambda}}
{(B-A)\tan\left(\sqrt{-\Lambda/3}\beta^{3/2}\right)-C/\sqrt{-3\Lambda}}\right|,
\end{equation}
where $A=\frac{m}{-2\Lambda}$ and
$B=\sqrt{\frac{m^2}{4\Lambda^2}+\frac{C^2}{-3\Lambda}}$ with the
scale factors being obtained from (\ref{W}). A glance at equations
(\ref{AO})-(\ref{AQ}) shows their regular behavior both in
Euclidean and Lorentzian regions passing smoothly through
signature changing hypersurface $\beta=0$. We should note that to
avoid imaginary values for $u(\beta)$ in (\ref{AO}), we must
restrict the allowed values of $\beta$ to the interval $0\leq
\beta \leq \beta_0=(\frac{3\pi^2}{-4\Lambda})^{1/3}$. However,
this is not a sever restriction since $\beta=0$ is the beginning
of the Lorentzian region and with the present bound on the
cosmological constant $|\Lambda|\sim 10^{-56}$ $cm^{-2}$, one has
$\beta_0\geq${\it present age of the universe}. In the limits of
$\beta\rightarrow 0$ and $\beta\rightarrow \beta_0$, equations
(\ref{AP}) and (\ref{AQ}) show that $v(\beta),w(\beta)\rightarrow
0$, {\it i.e.} the scale factors $a$, $b$ and $c$ become equal and
the universe approaches that of the flat Robertson-Walker in these
limits. In the case of $\Lambda>0$ the solutions can easily be
obtained by the replacement of the ``$\cos$'' function in
(\ref{AN}) with its hyperbolic counterpart and a quick look at the
resulting solutions shows that they do not exhibit the signature
changing behavior, see  \cite{16} for details.

\begin{figure}
\centerline{\begin{tabular}{ccccc}
\epsfig{figure=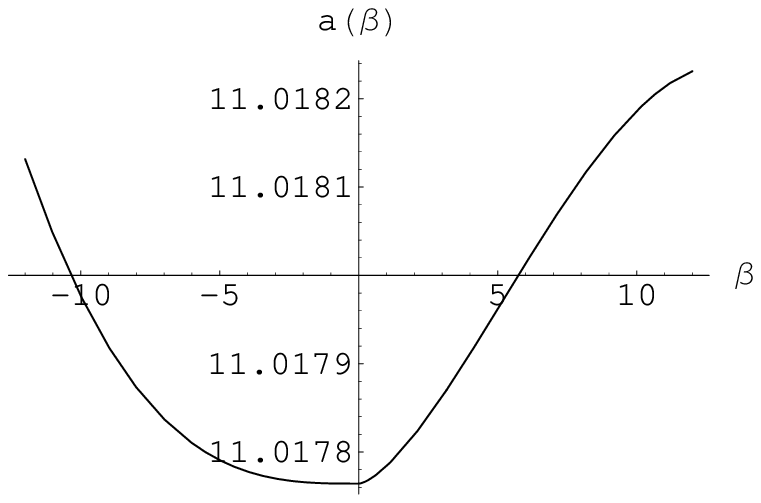,width=5cm}
 &\hspace{2.cm}&
\epsfig{figure=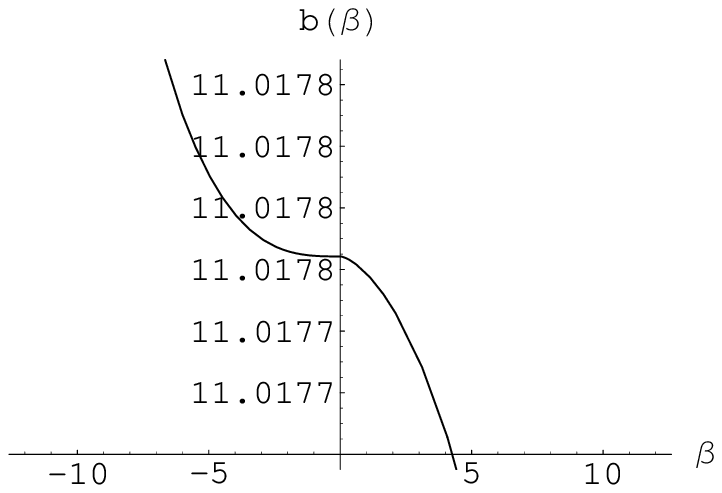,width=5cm}
 &\hspace{2.cm}&
\epsfig{figure=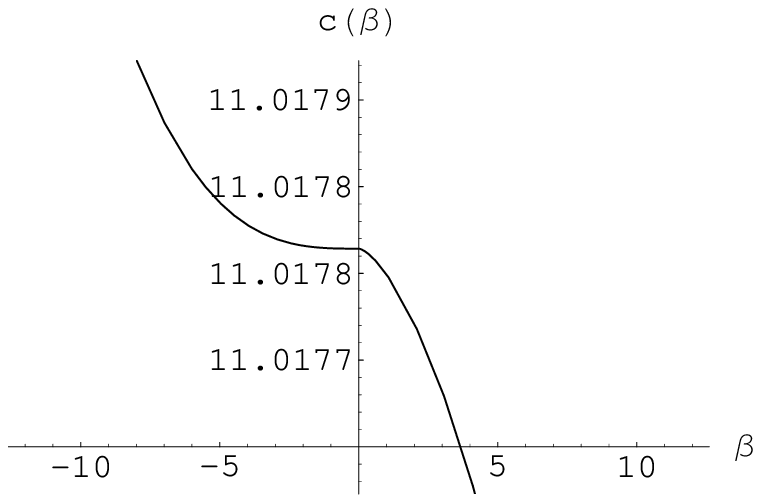,width=5cm}
\end{tabular}  }
\caption{\footnotesize The scale factors $a(\beta)$, $b(\beta)$
and $c(\beta)$ as functions of $\beta$ for $\Lambda=-10^{-6}$. The
values of other constants have been taken as 1.} \label{fig1}
\end{figure}

Thus, in summary, the above discussion shows that within the
context of our model, a universe with negative cosmological
constant can undergo signature transition from a Euclidean to a
Lorentzian domain through the $\beta=0$ hypersurface, where its
matter source is a spinor field. Indeed solutions
(\ref{AO})-(\ref{AQ}) describe the classical evolution of a
universe from a classically forbidden region where there is no
time, to a classically allowed Lorentzian region. This issue will
become more clear when we study the quantum cosmology of the model
in the next section. Figure 1 shows the behavior of the scale
factors $a(\beta)$, $b(\beta)$ and $c(\beta)$ as $\beta$ changes
from negative to positive values, showing a change of signature.
\section{Quantum cosmology}
The study of quantum cosmology of the model presented above is the
goal we shall pursue in this section. For this purpose we quantize
the dynamical variables of the model with the use of the WD
equation, that is, ${\cal H}\Psi=0$, where ${\cal H}$ is the
operator form of the Hamiltonian given by equation (\ref{AB}) and
$\Psi$ is the wave function of the universe, a function of the
scale factors and matter fields. With the replacement
$p_u\rightarrow -i\frac{\partial}{\partial u}$ and similarly for
$p_v$ and $p_w$, and choosing potential (\ref{AF}), the WD
equation reads
\begin{equation}\label{AR}
\left[\left(-\frac{\partial^2}{\partial
u^2}+\frac{\partial^2}{\partial v^2}+\frac{\partial^2}{\partial
w^2}\right)+12\left(-me^{3u}-\Lambda
e^{6u}+\lambda\right)\right]\Psi(u,v,w)=0,
\end{equation}
where we have used equation (\ref{AG}) to reduce the degrees of
freedom in the wave function. The solutions of the above
differential equation are separable in the form
$\Psi(u,v,w)=U(u)V(v)W(w)$, leading to
\begin{equation}\label{AS}
\frac{1}{V}\frac{d^2 V}{dv^2}=\pm \eta^2,
\end{equation}
\begin{equation}\label{AT}
\frac{1}{W}\frac{d^2 W}{dw^2}=\pm \zeta^2, \end{equation}
\begin{equation}\label{AU}
-\frac{d^2 U}{du^2}+12\left(-me^{3u}-\Lambda e^{6u}+\lambda \pm
\varsigma^2\right)U=0,
\end{equation}
where $\eta^2$ and $\zeta^2$ are separation constants and
$\varsigma^2=\frac{\eta^2+\zeta^2}{12}$. Before presenting the
solutions of the above equations, note that equation (\ref{AU}) is
a Schr\"{o}dinger-like equation for a particle with zero energy
moving in the field of the potential
\begin{equation}\label{AV}
{\cal U}(u)=-\Lambda e^{6u}-me^{3u}+\lambda \pm \varsigma^2.
\end{equation}
For a negative cosmological constant this potential has a minimum
at $u=\ln(\frac{m}{-2\Lambda})^{1/3}$. In the presence of this
potential the mini-superspace can be divided into two regions,
${\cal U}>0$ and ${\cal U}<0$ which could be termed as the
classically forbidden or Euclidean and classically allowed or
Lorentzian regions respectively. The boundary between the two
regions is given by ${\cal U}=0$, that is at
\begin{equation}\label{AW}
u_{\pm}=\ln \left[\frac{m}{-2\Lambda}\pm
\sqrt{\frac{m^2}{4\Lambda^2}-\frac{\lambda \pm
\varsigma^2}{-\Lambda}}\right]^{1/3}.
\end{equation}
Comparison of the above values of $u$ and the classical solutions
(\ref{AO}) suggests  that they correspond to $\beta=0$ and
$\beta=\beta_0$. Thus, the same boundary separates the Euclidean
and Lorentzian regions in both classical and quantum solutions. In
the Euclidean domain we have the wave-functions with exponential
behavior and in the Lorentzean region we have the wave-functions
of oscillatory nature \cite{3,4}. From equations
(\ref{AO})-(\ref{AQ}) it is clear that we have no classical
solutions that extend to infinite values of $u$, $v$ and $w$. Also
the beginning of evolution in the Lorentzian region is represented
by $v=w=0$. Thus we are led to the following boundary conditions
\begin{equation}\label{BB}
\begin{array}{c}
\Psi\rightarrow 0,\hspace{.5cm} \mbox{as} \hspace{.5cm} u,v,w\rightarrow +\infty, \\
\hspace{-.4cm}\Psi\rightarrow 0,\hspace{.5cm} \mbox{as} \hspace{.5cm} v,w\rightarrow -\infty. \\
\end{array}\end{equation}

The above discussion  suggests that we take the right hand side of
equations (\ref{AS}) and (\ref{AT}) with the upper sign, leading
to solutions
\begin{equation}\label{AX}
V(v)=e^{|\eta|v},\hspace{.5cm} W(w)=e^{|\zeta|w},\end{equation} in
the Euclidean region, and
\begin{equation}\label{AY}
V(v)=e^{-|\eta|v},\hspace{.5cm} W(w)=e^{-|\zeta|w},
\end{equation}
in the Lorentzian region. Now let us deal with the solution of
equation (\ref{AU}). This equation, after a change of variable
$\tau=\omega e^{3u}$ with $\omega=4\sqrt{-\Lambda/3}$ and
transformation $U=(\tau/\omega)^{-1/2}\phi$, becomes
\begin{equation}\label{AZ}
\frac{d^2 \phi}{d
\tau^2}+\left(\frac{-1}{4}+\frac{\kappa}{\tau}+\frac{1/4-\mu^2}{\tau^2}\right)\phi=0,
\end{equation}
where $\kappa=m/\sqrt{-3\Lambda}$ and
$\mu^2=\frac{4}{3}(\lambda+\varsigma^2)$. The above equation is
the Whittaker differential equation and its solutions can be
written in terms of confluent hypergeometric functions $M(a,b;x)$
and $U(a,b;x)$ as
\begin{equation}\label{BA}
\phi(\tau)=e^{-\tau/2}\tau^{\mu+1/2}\left[cU(\mu-\kappa+1/2,
2\mu+1;\tau)+c'M(\mu-\kappa+1/2,
2\mu+1;\tau)\right].
\end{equation}
Since the asymptotic behavior of $M(a,b;x)\sim e^x/x^{b-a}$
\cite{29}, we take $c'=0$. Therefore
\begin{equation}\label{BC}
U(\tau)=e^{-\tau/2}\tau^\mu U\left(\frac{1}{2}-\frac{4m}{3\omega}+
2\sqrt{\frac{\lambda+\varsigma^2}{3}},1+4\sqrt{\frac{\lambda+\varsigma^2}{3}};
\tau\right).
\end{equation}
A feature of the quantum cosmology of our model is the
quantization condition for the mass of the spinor field. A glance
at equations (\ref{AZ}) shows that its solutions have a behavior
of the form $\phi\sim e^{-\tau/2}$ when $\tau\rightarrow +\infty$
(or $u\rightarrow +\infty$) and $\phi\sim \tau^{\mu}$ in the limit
$\tau\rightarrow 0$ (or $u\rightarrow -\infty$). Demanding the
same limiting behavior for solution (\ref{BC}), one can easily see
that the function $U(a,b;x)$ should reduce to a polynomial and
this happens when $a=-n$ \cite{29}. This yields a quantization
condition for the mass of the spinor field as follows
\begin{equation}\label{BF}
m=\frac{3}{8}\left(2n+1+4\sqrt{\frac{\lambda+\varsigma^2}{3}}\right)\omega.
\end{equation}
Thus, the above discussions lead us to the following
eigenfunctions
\begin{equation}\label{BD}
\Psi_n(u,v,w)=e^{\pm|\eta|v\pm|\zeta|w}e^{-\omega
e^{3u}/2}e^{3\sqrt{(\lambda+\varsigma^2)/3}u}U\left(-n,
1+4\sqrt{\frac{\lambda+\varsigma^2}{3}};\omega e^{3u}\right),
\end{equation}
with the upper and lower signs signifying the Euclidean and
Lorentzian regions respectively. The general solution of the WD
equation can then written as
\begin{equation}\label{BM}
\Psi(u,v,w)=\sum_n c_n \Psi_n(u,v,w).
\end{equation}
To have an exponential wave function in the Euclidean domain we
must take $n=0$ in (\ref{BM}), {\it i.e.} the matter must be in
its ground state. Taking more terms in (\ref{BM}) leads us to the
wave function in the Lorentzian region. Summarizing, we have the
following wave functions
\begin{equation}\label{BN}
\Psi_E(u,v,w)=e^{|\eta|v+|\zeta|w}e^{-\omega
e^{3u}/2}e^{3\sqrt{(\lambda+\varsigma^2)/3}u}U\left(0,
1+4\sqrt{\frac{\lambda+\varsigma^2}{3}};\omega
e^{3u}\right),\end{equation} in the Euclidean region and
\begin{equation}\label{BE}
\Psi_L(u,v,w)=\sum_{n=1}^{\infty}c_n
e^{-|\eta|v-|\zeta|w}e^{-\omega
e^{3u}/2}e^{3\sqrt{(\lambda+\varsigma^2)/3}u}U\left(-n,
1+4\sqrt{\frac{\lambda+\varsigma^2}{3}};\omega e^{3u}\right),
\end{equation}
in the Lorentzian region. The wave functions (\ref{BN}) and
(\ref{BE}) describe a universe emerging out of the Euclidean
region with a smoothly changing signature and correspond to the
signature changing classical solutions. The creation of the
Lorentzian universe in this scenario which is characterized  by
the smooth passage through the Euclidean region is comparable to
the quantum tunneling from {\it nothing} in the Vilenkin's
proposal \cite{2}-\cite{6} where nothing is a 3-geometry of
vanishing size or a point. However, one should note that potential
(\ref{AV}) has no maximum and therefore should not be considered
as a potential barrier like those described in \cite{2}-\cite{6}.
Since the potential (\ref{AV}) has a dip the wave-function
(\ref{BE}) in the Lorentzian region resembles  bound states of the
Schr\"{o}dinger equation in a potential well.

In the case of a positive cosmological constant, potential
(\ref{AV}) is a monotonically  decreasing function of $u$.
Although in this case the mini-superspace can be divided into two
regions characterized by ${\cal U}>0$ and ${\cal U}<0$ too, in the
Lorentzian region, ${\cal U}<0$, we have solutions which extend to
infinite values of $u$, instead of being oscillatory as in the
case of $\Lambda<0$. Thus, there is no mechanism for a smooth
transition from the classically forbidden to the classically
allowed regions and the creation of a Lorentzian universe would
not be possible. We have seen before in section 4 that the
classical solutions with $\Lambda>0$ do not exhibit any signature
changing properties either, that is, the classical and quantum
solutions are in agreement in this case as well.
\section{Conclusions}
In this paper we have studied the classical and quantum evolution
of the cosmological solutions of the Einstein-Dirac system in a
Bianchi type I background. Our approach is a dynamical system
approach in which the mini-superspace is constructed from three
scale factors and components of the spinor field as the matter
source. From the classical solutions of this system we have chosen
those  that admit a degenerate metric for which the scale factors
of the universe have a continuous behavior in passing from a
classically forbidden (Euclidean) to a classically allowed
(Lorentzian) region. We have shown that this happens when the
cosmological constant is negative. The corresponding cosmology
begins in the Lorentzian domain by admitting a flat
Robertson-Walker metric, evolves according to equations
(\ref{AO})-(\ref{AQ}) and finally approaches a flat
Robertson-Walker universe again. The quantum cosmology of the
model presented above and the ensuing WD equation is amenable to
exact solutions in terms of confluent hypergeometric functions. We
have found that the division of the mini-superspace into the
classically forbidden and allowed regions also happens at the
quantum level. In the case of a negative cosmological constant the
behavior of the wave-functions are exponential in the Euclidean
domain and oscillatory in the Lorentzian region. These
wave-functions correspond to the classical signature changing
solutions and as such, could be useful in understanding the
initial conditions of the universe. Within the context of this
model the creation of the universe is described by a smooth
transition from a Euclidean to a Lorentzian region. Another
feature of the quantum cosmology of our model is a quantization
condition which leads to a spectrum for the allowed values of the
mass of the spinor field.

\end{document}